\def\c60{{\rm C}_{60}}
\def\Tc{T$_c$}

\def\etal{{\it et. al.}}

\def\Oh{{\hat \Omega}}

\def\bla{{{\bar \lambda}}}
\def\bmu{{{\bar \mu}}}
\def\be{\begin{equation}}
\def\ee{\end{equation}}
\def\bea{\begin{eqnarray}}
\def\eea{\end{eqnarray}}
\def\c60{C$_{60}$}
\def\bb#1{C$_{60}^{#1}$}
\def\a3c60{A$_3$C$_{60}$}

\def\bS{{\bf S}}
\def\bI{{\bf I}}

\def\cH{{\cal H}}

\def\cO{{\cal O}}
\def\half{{1\over 2}}

\def\cHt{{\tilde\cH}}

\def\bdper{\bfdelta_\perp}
\def\tpar{t^\parallel}
\def\tper{t^\perp}
\def\ubar{{\bar u}}
\def\yds#1{^\dagger_{#1}}
\def\uar{\uparrow}
\def\dar{\downarrow}

\def\kF{k_\ssr{F}}
\def\c60{~C$_{60}$}
\def\be{\begin{equation}}
\def\ee{\end{equation}}
\def\bea{\begin{eqnarray}}
\def\eea{\end{eqnarray}}
\def\bq{{\vec q}}
\def\br{{\bf r}}
\def\Oh{{\hat \Omega}}
\def\half{\frac{1}{2}}

  \font\elevenmib=cmmib10 scaled 1095
  \font\tenmib=cmmib10
  \font\eightmib=cmmib10 scaled 800
  \font\sixmib=cmmib10 scaled 667
  \skewchar\elevenmib='177
  \newfam\mibfam
  \def\mib{\fam\mibfam\tenmib}
  \textfont\mibfam=\tenmib
  \scriptfont\mibfam=\eightmib
  \scriptscriptfont\mibfam=\sixmib
  \mathchardef\alpha="710B
  \mathchardef\beta="710C
  \mathchardef\gamma="710D
  \mathchardef\delta="710E
  \mathchardef\epsilon="710F
  \mathchardef\zeta="7110
  \mathchardef\eta="7111
  \mathchardef\theta="7112
  \mathchardef\kappa="7114
  \mathchardef\lambda="7115
  \mathchardef\mu="7116
  \mathchardef\nu="7117
  \mathchardef\xi="7118
  \mathchardef\pi="7119
  \mathchardef\rho="711A
  \mathchardef\sigma="711B
  \mathchardef\tau="711C
  \mathchardef\phi="711E
  \mathchardef\chi="711F
  \mathchardef\psi="7120
  \mathchardef\omega="7121
  \mathchardef\varepsilon="7122
  \mathchardef\vartheta="7123
  \mathchardef\varrho="7125
  \mathchardef\varphi="7127

\def\etal{{\it et al.\/}}

\def\ie{{\it i.e.\/}}

\def\eg{{\it e.g.\/}}

\def\sss#1{{\scriptscriptstyle #1}}

\def\ssr#1{{\sss{\rm #1}}}

\def\dsl{\raise.15ex\hbox{$/$}\kern-.57em\hbox{$\partial$}}
\def\nsl{\raise.15ex\hbox{$/$}\kern-.57em\hbox{$\nabla$}}
\def\id{\raise.72ex\hbox{$-$}\kern-.85em\hbox{$d$}\,}
\def\gtwid{\,{\raise.3ex\hbox{$>$\kern-.75em\lower1ex\hbox{$\sim$}}}\,}
\def\ltwid{\,{\raise.3ex\hbox{$<$\kern-.75em\lower1ex\hbox{$\sim$}}}\,}
\def\undr{\raise.3ex\hbox{$\sim$\kern-.75em\lower1ex\hbox{$|\vec x|\to\infty$}}}

\def\frac#1#2{{\textstyle{#1 \over #2}}}
\def\half{\frac{1}{2}}

\def\fourth{\frac{1}{4}}

\def\sbl{\left [}
\def\sbr{\right ]}
\def\({\left (}
\def\){\right )}

\def\zhat{{\hat{\bf z}}}

\def\cH{{\cal H}}

\def\cO{{\cal O}}

\def\bfI{{\mib I}}

\def\bfS{{\mib S}}

\def\bfc{{\mib c}}

\def\bfi{{\mib i}}

\def\bfq{{\mib q}}

\def\bfdelta{{\mib\delta}}

\def\bftau{{\mib\tau}}

\def\xhi{{\raise.35ex\hbox{$\chi$}}}

\def\rmA{{\rm A}}

\def\rmC{{\rm C}}

\def\rmF{{\rm F}}

\def\rmN{{\rm N}}

\def\ket#1{{\,|\,#1\,\rangle\,}}
\def\bra#1{{\,\langle\,#1\,|\,}}

\def\yds#1{^{\dagger #1}}
\def\nds#1{_{#1}^{\vphantom{\dagger}}}

\def\and{a^{\phantom\dagger}}

\gdef\journal#1, #2, #3, 1#4#5#6{               
    {\sl #1~}{\bf #2}, #3 (1#4#5#6)}            
\def\pr{\journal Phys. Rev., }

\def\prb{\journal Phys. Rev. B, }

\def\prl{\journal Phys. Rev. Lett., }

\def\nupb{\journal Nucl. Phys. B, }

\def\phyla{\journal Phys. Lett. A, }

\def\jmp{\journal J. Math. Phys., }

\documentstyle[12pt]{article} 
\begin{document} 
\baselineskip 18pt
\title{Electronic Properties of Buckminsterfullerenes: Degeneracies and Surprises}
\author{Assa Auerbach \\ Physics
Department, Technion-IIT, Haifa 32000, Israel\thanks{Email: assa@pharaoh.technion.ac.il}.}
\maketitle 
\begin{abstract} 
Buckminsterfullerene compounds exibit remarkable physics at low temperatures, e.g.
high temperature superconductivity  in alkali\--full\--erenes,
and ferromagnetism  in TDAE\--\c60.  Here we review recent theoretical studies of electron correlations in these compounds. In particular, we discuss models of electron-vibron interactions, electron-electron interactions, and intermolecular hopping.
We show that the origin
of novel electronic phases lies in {\em local}  degeneracies of
\c60; a direct consequence of the high molecular symmetry.
\end{abstract}
 
\section{Introduction}

The synthesis of buckminsterfullerene (\c60) into molecular crystals
with electron donors has resulted in materials with surprising
electronic properties\cite{hebard}. 

C$_{60}$ is a truncated
icosahedron. From a physicist's standpoint, the charged molecule is
fundamentally interesting, because the high molecular symmetry gives rise to
degeneracies in both electronic and vibrational systems. Thus, the  molecule is
very sensitive to perturbations. In particular, electron--phonon and
electron--electron interactions are expected to produce highly correlated
ground states and excitations.

First, superconductivity was
discovered in alkali-fullerenes, \a3c60, (A=K, Cs, Rb)  at
relatively high temperatures 
(\Tc$\le 33^\circ$K \cite{c60}), compared, say,  to  intercalated graphite.
Soon thereafter ferromagnetism was found in TDAE-\c60 at
 \Tc$\approx 16^\circ$K \cite{tdae-fm,tdae-fm1,rao1}, where TDAE is
tetr\-akis\-(di\-methyl\-amino)\-ethy\-lene, C$_2$N$_4$(CH$_3$)$_8$ (the
stoichiometry is a 1:1 ratio of
TDAE to \c60).  

Some striking aspects of  TDAE-\c60  are its relatively large
value of  \Tc  -- for an organic ferromagnet -- and
and its nonmetallic conductivity, suggestive of Mott-Hubbard
localization\cite{microwave}.  Ferromagnetism is thus particularly unusual, since
superexchange na{\"\i}vely predicts {\em   antiferromagnetic} interactions
between localized spins in many such systems.

An obvious question is raised: 
{\it Are superconductivity in \a3c60 and ferromagnetism
in TDAE-\c60  related?}  While superconductivity involves effectively
attractive interactions, magnetism is usually
believed to result from repulsive Coulomb
forces.\footnote{A similar mystery underlies the
proximity of high  \Tc  superconductivity to antiferromagnetism in
the high- \Tc  cuprates.} 
The similarity between the materials is that both involve
partially filled conduction bands made of  $t_{1u}$-orbitals of \c60.
Also one expects  similar  intramolecular electron-vibron and
electron-electron interactions.

The primary differences between the two systems are in their crystalline
symmetries and \c60 ionizations.
While \a3c60 is an FCC crystal with cubic symmetry in which the
\c60 molecules are triply ionized (\bb{3-}),
TDAE-\c60 has a $c$-centered monoclinic unit cell\cite{structure},
which gives rise to preferred hopping along the $c$-axis,
and singly ionized \bb{-}.

Here we  review recent work on the local molecular
interactions which are separated into two major contributions: 
the electron-electron pseudopotentials (Section \ref{sec2}) , and the electron-vibron interactions (Section \ref{sec3}).  The
latter  exhibit, in the semiclassical limit, interesting dynamical Jahn Teller effects accompanied by non trivial Berry phases \cite{tosatti,assa-mab,amt}. We show that this effect enhances the attractive interaction due to electron-vibron interactions
by a significant factor. This analysis,  though admittedly qualitative,
points at a plausible cause for the relatively high \Tc's of  
\a3c60.

In Section \ref{sec4} we discuss TDAE-\c60. We derive an effective model for
the low excitations of  TDAE-\c60 using a muticomponent superexchange expansion
about the  Mott-Hubbard insulating phase \cite{assa-dan}. 
As a result, we obtain a rich phase diagram
which includes spin ferromagnetism, spin density waves, and orbital
ferromagnetism depending on
the ratios of interaction pseudopotentials.

\section{The electron-electron pseudopotentials}
\label{sec2}
We consider a system of  $N$ $\pi$-electrons
hopping on the truncated icosahedron (the soccer-ball) lattice of \c60.
The single particle 
Hamiltonian is 
\be
H^0~=
\sum_{\lambda} \epsilon_\lambda
\sum_{\mu} ~\sum_{s=\uparrow\downarrow}
 ~c^\dagger_{\lambda\mu s} c_{\lambda\mu s}~
\label{h0}
\ee
$\lambda\mu$ denote the irreducible representations of the icosahedral group $\chi_{\lambda\mu }(\Oh)$
where $\Oh$ is a unit vector pointing at one of the
soccer ball vertices. At low $\lambda$, $\chi_{\lambda\mu }(\Oh)$ are 
closely related to
the angular momentum functions  $Y_{lm}(\Oh)$.
However, the structure of
the states near the Fermi level of 60 electrons is different. The
soccer-ball lattice splits
the degenerate spherical $l=5$ multiplet, and the Fermi level lies in
the semiconducting gap between the 5-fold and 3-fold degenerate
$\lambda=h_g, t_{1u}$ orbitals respectively.

The  Coulomb interactions on a metallic spherical shell is
parametrized by two dimensionless variables
$g,\alpha$.
\bea
H^{e-e} &=&\half \int d\Oh_1 d\Oh_2 V 
(\Oh_1-\Oh_2)~:\rho(\Oh_1)\rho(\Oh_2): \nonumber\\
V &=&
g ~2 \pi r_{min}\sum_{LMss'} \left({ 2 \over r_{min} ( 2L+1) }
\right)^\alpha ~Y^*_{LM}(\Oh_1) Y_{LM}(\Oh_2)~~,
\label{int}
\eea
$\rho=\sum_s \psi^\dagger_s\psi_s$ is the density operator where
$\psi^\dagger_s ~= \sum_{\lambda\mu} \chi_{\lambda\mu}(\Oh) c^\dagger_{\lambda\mu s}$.
$g~= e^2/(\epsilon R e_0)$ is the {\it strength} of the 
screened interaction, and $R$ is the radius of the ball.
The power law tail of
$V(r)$ depends on $\alpha$ such that as $\alpha$ decreases,  the
potential has shorter range. $0\le \alpha
\le 1$
interpolates smoothly between a
$\delta$ function and an  unscreened $1/r$ Coulomb potential:
\be
V^{\alpha=0}~= g ~ 2\pi r_{min}  \delta (\Oh_1 - \Oh_2)~~;~~
V^{\alpha=1}~= {g\over |\Oh_1 - \Oh_2|}
\ee
When $\alpha$ changes from 0 to 1, the averaged
interaction 
over an area of radius $r_{min}$ (analogous to Hubbard's U) is held fixed.

The interaction in the basis of $H_0$ can thus be explicitly given in the second quantized form as
\bea
H^{e-e}&=&g\pi 
r_{min}\sum_L \left( {2\over r_{min}(2L+1)
}\right)^\alpha \nonumber\\
&&~~~\times (-1)^{m'+\mu}~C^{L~\bla~\bla'}_{M \bmu ~-\bmu'}~C^{L\lambda ~\lambda'}_{M
\mu~-\mu'} 
c^\dagger_{\bla'\bmu's}c^\dagger_{\lambda'\mu'\sigma}
c_{\lambda\mu\sigma}c_{\bla\bmu s}\nonumber\\
C^{L~\lambda~\lambda'}_{M \mu ~-\mu'}&\equiv & {1\over 4\pi} \int d\Oh Y^*_{LM}(\Oh) \chi^*_{\lambda\mu}(\Oh)\chi_{\lambda'\mu'}(\Oh)
\eea
where summation of repeated indices is assumed.

For a molecule with $N$ electrons in a closed shell (above which there is a gap larger than the interaction strength),  the zero frequency interaction between two electrons  in the open shell
is given by
\be
u_{\lambda\mu}~\equiv~ E_{N+2}^{\lambda\mu}-~2E_{N+1}+~E_N 
\label{uL}
\ee 
which we call ``pseudopotentials''. 
 ${\lambda\mu}$ denote  the representation of the $N+2$ electron state. 
The computation of $u_{\lambda\mu}$ for the Hamiltonian  $H^0+H^{e-e}$
is a difficult task which was undertaken using second order perturbation theory\cite{ck,assa-gan} and later using perturbative renormalization group up to three loops order\cite{Murthy}. 

For the real compounds however, estimation of the
true pseudopotentials requires incorporating the electron-vibron interactions (see the following Section \ref{sec3}) as well as crystal fields, screening from neighboring molecules and polarization effects of $\sigma$ electrons. As we shall see, the electronic properties of the conduction bands in the solid are very sensitive to
the combination of electron-electron and electron-vibron
pseudopotentials. Thus even ab-initio
calculations\cite{gunnar} (which involve certain approximations for
the higher order effects of interactions) may not be  accurate enough for precise predictions of the ground state phase diagram.  Here we shall therefore look for
{\em qualitative} effects and leave the electron-electron pseudopotentials as free parameters within an acceptable range of magnitudes.

\section{The electron-vibron problem and superconductivity}
\label{sec3}
In this section  we solve  the problem of a single $H_g$ vibron
coupled to $t_{1u}$ electrons in a C$_{60}^{n-}$ molecule. The model
is too simplified for quantitative predictions for C$_{60}$, but it contains
interesting  novel physics which will be important for further
studies of this system.

Semiclassically, a dynamical Jahn--Teller effect occurs\cite{jt,vzr,t1}.  For $n=1,2,4,5$, the
molecule distorts unimodally, giving rise to a pseudo-angular momentum
spectrum, plus three harmonic oscillators. For $n\!=\!3$, there is a bimodal
distortion, which generates a spectrum of a symmetric top rotator, plus two
harmonic oscillators. The pseudo--rotations are subject to non trivial Berry
phase effects\cite{lh}, which determine the  pseudo-angular momenta $L$, and thus the
degeneracies and level ordering of the low lying states. Strong Berry phase
effects seem to survive even at moderate and weak coupling as shown by the
exact diagonalization results.

We find at weak coupling that the pair binding energy is a factor of $5/2$
larger than the classical JT effect, and a factor of three larger
than the pairing interaction of  Migdal--Eliashberg theory of
superconductivity. This enhancement can be interpreted semiclassically as due
to large zero point energy reduction of the pseudo--rotations. From the weak
coupling point of view, this effect is due to degeneracies in both electronic
and vibronic systems.

Migdal's approximation neglects vertex corrections in the resummation of
two--particle ladder diagrams. This is justified only in the retarded limit
$\omega << \epsilon_F$. Here we have considered the  opposite limit, where the
molecular ground state energies are solved first, assuming that the JT
relaxation time is  of the same order, or faster than the inter molecular
hopping time. In this regime, we have found therefore that Migdal's
approximation  substantially {\em underestimates} the pairing interaction, and
$T_c$, for these ideal molecular solids \cite{pietron}. This large effect
suggests that some of the enhancement is likely to carry over to the real case
of A$_3$C$_{60}$ metals, where electron hopping $t$ and vibron frequencies are
of similar strength.
\subsection{The Model}

We consider a single  $H_g$  (five
dimensional) vibrational multiplet which couples to $n=1,\ldots 5$ electrons in an open 
$t_{1u}$ shell.  $t_{1u}$ and $H_g$ are the icosahedral group counterparts of the
spherical harmonics  $\{Y_{1m}\}_{m=-1}^1$, and  $\{Y_{2M}\}_{M=-2}^2 $
respectively. By replacing  the truncated icosahedron (soccer ball) symmetry
group by the spherical group, we ignore lattice corrugation effects. These are
expected to be small since they do not lift
the  degeneracies of the $L=1,2$ representations.

The  Hamiltonian is thus defined as\cite{amt}
\be
H~=H^0 ~+ H^{e-v} ~,
\label{2.1}
\ee
where,
\be
H^0=  \hbar \omega \sum_M \left
(b^\dagger_{ M } b_{M}+\half\right)~+(\epsilon-\mu)\sum_{ms}
c^\dagger_{ms} c_{ms} ~.
\label{2.2}
\ee
$b^\dagger_{M}$ creates a vibron with azimuthal quantum number $M$,  and
$c^\dagger_{ms}$ creates an electron of spin $s$ in an orbital $Y_{1m}$.
By setting $\mu\to \epsilon$ we can discard the second term.

The $H_g$  vibration field is
\be
 u  (\Oh) =   {1\over \sqrt{2}} (Y^*_{2M}(\Oh) b^\dagger_{ M }+Y_{2M}(\Oh)
b_{M}) ~,
\label{2.3}
\ee
where $\Oh$
is a unit vector on the sphere.  The $t_{1u}$ electron field is
 \be
\psi^\dagger_s(\Oh)=\sum_{m=-1}^1 Y_{1m}(\Oh) c^\dagger_{ms} ~.
\label{2.4}
\ee
The  electron--vibron interaction  is local and rotationally invariant.
Its form is
completely determined (up to an overall coupling constant $g$) by
symmetry:
\be
H^{e-v}~\propto   g \int\! d\Oh u (\Oh) \sum_s
\psi^\dagger_s(\Oh)\psi_s(\Oh) ~.
\label{2.5}
\ee
Using the relation
\be
\int \!d\Oh ~Y_{LM}(\Oh ) Y_{lm_1}(\Oh) Y_{lm_2}(\Oh) \propto (-1)^M \langle
L,-M|lm_1;lm_2\rangle ~,
\label{2.6}
\ee
where
$\langle \cdots \rangle$ is a Clebsch-Gordan coefficient \cite{ed},
yields the  second quantized Hamiltonian
\bea
H^{e-v}&=&  {\sqrt{3}\over 2} g \hbar \omega \sum_{s,M,m}
(-1)^{m} \left(b^\dagger_M+(-1)^M b_{-M}\right)
\nonumber\\ &&~~~~~~~~~~~~~~~\times  \langle 2, M|1,-m;1,M+m\rangle
c^\dagger_{m  s }c_{M+m s} ~.
\label{2.7}
\eea
The coupling constant $g$ is fixed by the convention of O'Brien, who studied
first this kind of dynamical JT problem \cite{ob1}.  Representation (\ref{2.7})
is convenient for  setting up an exact diagonalization program in the truncated
Fock space.

\subsection{The Real Representation}
The semiclassical expansion  is simpler to derive in the real coordinates
representation. The vibron  coordinates are
 \be
q_{\mu}~={6\over\sqrt{10}}
 \sum_{m=-2}^2 M_{\mu m}  \left(b^\dagger_{m}+(-1)^{m} b_{-m}\right) ~,
\ee
where
\bea
M_{\mu,m \ne 0}&=& (2~ {\rm sign} (\mu) )^{-\frac{1}{2}}
 \left(\delta_{\mu,m} +
 {\rm sign} (\mu)\delta_{\mu,-m}\right),\nonumber\\
 M_{\mu,0}&=& \delta_{\mu,0}.
\label{2.8.0}
\eea
$\{q_\mu\}$ are coefficients of the real spherical functions
\bea
f_\mu(\Oh) &=& {6\over\sqrt{5}} \sum_m M_{\mu,m} Y_{2 m}(\Oh)\nonumber\\
&=& \cases{{6\over\sqrt{10}}~Re\left(Y_{2|\mu|}(\Oh)\right)  & $\mu = 1,2$\cr
 {6\over\sqrt{5}}~ Y_{20}(\Oh)  & $\mu = 0$\cr
{6\over\sqrt{10}}~Im \left(Y_{2|\mu|}(\Oh)\right) & $\mu =-1,-2$}~.
\label{2.8}
\eea
We also choose a real representation for the  electrons
\bea
c^\dagger_{xs}&=& {1\over \sqrt{2}}\left( c^\dagger_{1s} +
c^\dagger_{-1s}\right)\nonumber\\
c^\dagger_{ys}&=& {1\over i\sqrt{2}}\left( c^\dagger_{1s} -
c^\dagger_{-1s}\right)\nonumber\\
c^\dagger_{zs}&=&   c^\dagger_{0s} ~.
\eea
Thus the Hamiltonian in the real representation is given by
\bea
H &=& H^0~+H^{e-v}\nonumber \\
H^0 &=&{ \hbar \omega\over 2} \sum_{\mu }\left( -
\partial_\mu^2   +   q_{\mu}^2\right) \nonumber \\
H^{e-v} &=& g {\hbar\omega\over 2}  \sum_s
(c^\dagger_{xs}, c^\dagger_{ys},c^\dagger_{zs})
\pmatrix{
q_0+\sqrt{3}q_2	&-\sqrt{3}q_{-2}	&\sqrt{3}q_1 \cr
-\sqrt{3}q_{-2}	&q_{0}-\sqrt{3}q_{2}	&-\sqrt{3}q_{-1}\cr
-\sqrt{3}q_1	& \sqrt{3}q_{-1}	& -2q_{0}     }
					\pmatrix{	c_{xs}\cr
                        	                	c_{ys}\cr
							c_{zs} }\nonumber \\
\label{2.9}
\eea

This form of the JT hamiltonian is well known \cite{ob1,vzr}. Since the
Hamiltonian is rotationally invariant, its eigenvalues  are invariant under
simultaneous O(3) rotations of the electronic and vibronic representations.

\subsection{Jahn--Teller Distortions}
\label{class}
In the classical limit, one  can ignore  the vibron derivative terms in
(\ref{2.9}), and treat  $\bq=\{q_\mu\}$ as frozen coordinates in $H^{e-v}$. The
coupling matrix in $H^{e-v}$ is diagonalized by \cite{ob2}:
\be
  T^{-1}(\varpi)~ \pmatrix{z - \sqrt{3}r  &0&0\cr 0& z +
\sqrt{3}r   &0\cr 0&0&-2z } ~T(\varpi) ~,
\label{3.1}
\ee
where
\be
T~=   \pmatrix{ \cos \psi
& \sin \psi& 0\cr -\sin \psi& \cos \psi &0\cr
0&0&1}
  \pmatrix{ \cos\theta &0& \sin\theta\cr
0&1&0\cr
\sin\theta&0&\cos\theta}
  \pmatrix{ \cos\phi &  \sin\phi&0\cr
-\sin\phi& \cos\phi & 0\cr
0&0&1}.
\label{3.2}
\ee
$\varpi=(\phi,\theta,\psi)$  are the three Euler angles of the O(3) rotation
matrix $T$. In the diagonal basis of (\ref{3.1}),  the electron energies depend
only on  two  vibron coordinates:
\be
{\bq}(0)  =\pmatrix{r\cr
                          0\cr
		               z \cr
                          0\cr
                          0} .
\label{3.3}
\ee
By rotating
the vibron coordinates $\bq$ to the diagonal basis using  the  $L=2$
rotation  matrix $D^{(2)}$ \cite{ed},  one obtains
\be
{\bq}_\mu (r,z,\varpi)~= \sum_{m,m',\mu'=-2}^2 M_{\mu,m}
D^{(2)}_{m,m'}(\varpi) M_{m'\mu'}^{-1} {\bq}_{\mu'}(0) ~,
\label{3.3.1}
\ee
where $M_{\mu,m}$ was defined in (\ref{2.8.0}).

By (\ref{3.3.1}), and the unitarity of $D$ and $M$, $|{\bq}|^2$ is invariant
under rotations of $\varpi$. Thus, the adiabatic potential energy $V$ depends
only on $r$,  $z$, and the occupation numbers of  the electronic eigenstates $
n_i$, where  $\sum_i n_i=n$.
\be
V(z, r,[n_i]) = {\hbar\omega\over 2} ( z^2+r^2)~+  {\hbar\omega
g\over 2}\left( n_1(z-\sqrt{3}r) +n_2(z+\sqrt{3}r) - n_3 2 z \right)  .
\label{3.4}
\ee
$V$ is minimized at the JT distortions
$({\bar z}_n,{\bar
r}_n,{\bar n}_i)$, at which the classical energy is given by
\be
E_n^{cl} ~= \mbox{min}~ V({\bar z}_n,{\bar
r}_n,{\bar n}_i) .
\label{3.5}
\ee
The JT distortions at different fillings are given in Table I. 
We define
${\tilde \phi},{\tilde\theta}$ as the longitude and latitude with respect to
the diagonal frame (``principal axes'') labelled $(1,2,3)$ ($3$ is at the north
pole). ${\bar z},{\bar r}$ parametrize the Jahn-Teller distortion in the real
representation (\ref{2.3}), as
\be
\langle u^{JT}(\Oh)\rangle ~=
 { {\bar z}\over 2}(3\cos^2{\tilde\theta} -1 )~+
 {{\bar r} \sqrt{3}\over 2} \sin^2{\tilde\theta}\cos(2{\tilde\phi}).
\label{3.6}
\ee
In Table I we present the values of the ground state JT distortions at
all electron fillings. We see that electron fillings $n$ = 1, 2, 4, 5 have
{\em unimodal distortions } which are symmetric about the $3$ axis,  while $n=3
$ has a {\em bimodal}, about the $3$ and $1$ axes. The two types of
distortions are portrayed in Fig. \ref{dist}. We depict the distortions of
(\ref{3.6}) for the unimodal and bimodal cases.

\subsection{Semiclassical Quantization}
\label{semiclas}
At finite coupling constant $g$,  quantum fluctuations about the frozen JT
distortion must be included. In order to carry out the semiclassical
quantization, we define a  natural set of five dimensional coordinates
$r,z,\varpi$.  $\varpi$ parametrize the motion in the JT manifold (the valley
in the ``mexican hat'' potential $V$) and $r,z$ are  transverse to the JT
manifold, since $V$ depends on them explicitly.   

For the unimodal cases the kinetic energy is given by  
 \be
\frac{1}{2} |{\dot {\vec q}}|^2 \approx  \half \left( {\dot z}^2 + {\dot r}^2
+  r^2 (2{\dot \psi})^2 ~+  3{\bar z}^2\left( {\dot\theta}^2 + \sin^2\theta
{\dot \phi}^2 \right) \right) .
\label{4.6}
\ee
and the semiclassical Hamiltonian is thus
\bea
H^{uni} &\approx& H^{rot} + H^{ho}\nonumber\\
H^{rot} &=& {\hbar \omega \over 6{\bar z}^2} {\vec L}^2\nonumber\\
H^{ho}&=&{\hbar \omega} \sum_{\gamma=1}^3 (a^\dagger_\gamma a_\gamma + \half)
{}~,
\label{4.10}
\eea
where  ${\vec L}$ is an angular momentum operator, and $H^{ho}$ are the three
harmonic oscillator modes of $\br$.
The energies are given by
\be
E^{uni} ~= {\hbar \omega }\left({1\over 6{\bar z}_n^2 }L(L+1) +
 \sum_{\gamma=1}^3
(n_\gamma +\frac{1}{2} )\right) .
\label{4.10.1}
\ee
The rotational part of the eigenfunctions is
\be
\Psi^{rot}_{Lm} ({\bq}) =  Y_{Lm}(\Oh) ~  |[n_{is}]\rangle_\Oh ~,
\label{4.11}
\ee
where $\Oh=(\theta,\phi)$ is a unit vector, and $ |[n_{is}]\rangle_\Oh$ is the
electronic adiabatic ground state. It is a Fock state in the {\em principal
axes}  basis.  In terms of the stationary Fock basis $|[n_{\alpha s'}]\rangle$
where $\alpha=x,y,z$, the adiabatic ground state is
\be
|[n_{is}]\rangle_\Oh ~= \sum_{[n_{\alpha s }]}
   \langle[n_{\alpha s }] |[n_{is}]\rangle_\Oh
 |[n_{\alpha s }]\rangle .
\label{4.12}
\ee
Each overlap is a Slater determinant which is a sum of $n$ products of
spherical harmonics
\be
\langle[n_{\alpha s}] |[n_{is}]\rangle_\Oh ~=
\sum_{[\nu]} C_{[\nu]} Y_{1\nu_1}(\Oh) Y_{1\nu_2}(\Oh)\cdots Y_{1 \nu_n}(\Oh)
{}~,
\label{4.12.1}
\ee
where $C_{[\nu]}$ are constants.

Now we discuss how boundary conditions determine the allowed values of $L$.
A reflection on the JT manifold  is given by
\be
\Oh ~\to -\Oh .
\label{4.13}
\ee
Spherical harmonics are known to transform under reflection as
\be
Y_{L m} \to\!(-1)^LY_{L m} ~.
\ee
Thus, by (\ref{4.12}) and (\ref{4.12.1}), the electronic part of the wave
function transforms as
\be
|[n_{is}]\rangle_\Oh \to (-1)^n |[n_{is}]\rangle_{ -\Oh } ~.
\label{4.14}
\ee

The reflection (\ref{4.13}) can be performed by moving on a {\em  continuous
path} on the sphere from any point to its opposite.  
It is easy to verify, using (\ref{3.3.1}) that this path {\em
is a closed orbit of ${\vec  q} \in R^5$}:
\be
{\bq}(\Oh)\to {\bq}(-\Oh) ={\bq}(\Oh) ~.
\label{4.15}
\ee
Thus we find that the electronic wave function yields a {\em Berry phase
factor} of $(-1)^n$ for rotations between opposite points on the sphere  which
correspond to closed orbits of ${\bq}$. In order to satisfy (\ref{4.11}))
using  the invariance  of the left hand side under reflection, the
pseudorotational $Y_{Lm}$ wavefunction must cancel the electronic Berry phase.
This amounts to a {\em selection rule} on $L$:
\be
(-1)^{L+n} = 1 ~.
\label{4.16}
\ee
Thus, the ground state for $n = 1$ and $5$ electrons has pseudo-angular
momentum
$L = 1$ and finite zero point  energy due to the non trivial Berry phases.

\subsubsection{Bimodal Distortion}
The analysis of the bimodal distortions  $n = 3$ proceeds along similar lines.
The quantization of the pseudo--rotational part
 is the quantum symmetric top Hamiltonian. Fortunately, it is
a well-known textbook problem (see e.g. Ref. \cite{ll,ed}).
The eigenfunctions of a rigid body rotator are the rotational matrices
\be
D_{mk}^{(L)} (\varpi) ,
\ee
where $ L,m,k $ are quantum numbers of the commuting operators ${\vec L}^2,
L^z, L^1$ respectively. $L^z$ and $L^1$ are defined with respect to the fixed
$z$ axis and the co-rotating $1$ axis respectively. The quantum  numbers are in
the ranges
\bea
L&=& 0,1,\ldots \infty\nonumber\\
m,k &=& -L, -L+1,\ldots L ~.
\label{4.20.1}
\eea
The remaining coordinates are two massive
harmonic oscillators modes
\be
\br = (r-{\bar r},z-{\bar z}).
\label{4.21}
\ee
The semiclassical Hamiltonian is thus
\bea
H^{bi} &\approx& H^{rot} + H^{ho},\nonumber\\
H^{rot} &=& {\hbar \omega \over 4{\bar z}^2 } {\vec L}^2 -
{3\hbar \omega \over 16{\bar z}^2} (L^1)^2,\nonumber\\
H^{ho}&=&{\hbar \omega} \sum_{\gamma=1}^2 (a^\dagger_\gamma a_\gamma + \half),
\label{4.22}
\eea
and its eigenvalues are
\be
E^{bi}={\hbar \omega }\left({1\over 4{\bar z}^2  }L(L+1)-
 {3\over 16{\bar z}^2}k^2~+\sum_{\gamma=1}^2
(n_\gamma +\frac{1}{2} )\right) .
\label{4.23}
\ee
The rotational eigenfunctions are explicitly dependent on $\varpi$ as
\be
\Psi^{rot}_{Lmk}[{\bq}] =   D_{mk}^{(L)} (\varpi)
\prod_{is}|n_{is}\rangle_{\varpi} ~ .
\label{8.1}
\ee

Unlike the unimodal case, in the bimodal case no single reflection fully
classifies the symmetry of the wavefunction.  However, one can obtain definite
sign factors by transporting the electronic ground state  in certain orbits. We
define the rotations of $\pi$ about principle axis $L^i$  as $C_i$. 
The Berry phases associated with
these rotations can be read directly from the rotation matrix $T$ in  Eq.
(\ref{3.2}).  For example: for $\psi\!\to\!\psi+\pi$ ($C_3$), the states
$|1\rangle$ and $|2\rangle$ get  multiplied by $(-1)$.

Since $D^{(L)}_{m,k}$ transform  as $Y_{Lk}$ under $C_i$,  it is easy to
determine the  sign factors of the pseudorotational wavefunction. The results
are given below:
\begin{eqnarray}
C_1:|1,0,2\rangle_{\varpi}  \to |1,0,2\rangle_{\varpi'} &~~~& C_1:D^{(L)}_{m,k}
\to (-1)^k D^{(L)}_{m,k} \nonumber\\
C_2:|1,0,2\rangle_{\varpi}  \to -|1,0,2\rangle_{\varpi'} &~~~&
C_2:D^{(L)}_{m,k}
\to (-1)^{L+k} D^{(L)}_{m,-k}
\nonumber\\
C_3:|1,0,2\rangle_{\varpi}  \to -|1,0,2\rangle_{\varpi'} &~~~&
C_3:D^{(L)}_{m,k}
\to  (-1)^{L} D^{(L)}_{m,-k} ~.
\label{9}
\end{eqnarray}
${\bq}$ are coefficients in an $L=2$ representation, and therefore are
invariant under $C_1,C_2,C_3$. $C_i$ describe continuous closed orbits in
$R^5$.
In order to satisfy (\ref{9}) and using the degeneracy of $E^{bi}$ for
$k\!\to\!-k$,  we find that
\be
L~=~\mbox{odd}~,~~~~~k ~=~\mbox{even}.
\ee
In
particular, the ground state
of (\ref{8.1}) is given by
$L\!=\!1$, and $k\!=\!0$.

\subsection{Exact Diagonalization}
\label{exact}

The above semiclassical scheme gives a clear and intuitive picture of the
behaviour of the system in a strong coupling limit\cite{assa-mab}. This limit is appropriate
for describing, e.g., Na$_3$ \cite{delac}. However, in C$_{60}$ the actual
range of the coupling parameter - $g\approx 0.3$ for a typical mode
\cite{antr,II} - suggests that the electron--vibron coupling is actually in the
weak to intermediate regime.

In Ref. (\cite{amt}) we have diagonalized the electron--vibron Hamiltonian (\ref{2.7}) for single $H_g$ mode  in a truncated Fock space. This approach yields accurate results
weak to moderate values of the coupling strength.  The eigenenergies  match  the semiclassical approximation (\ref{4.10.1}) and (\ref{4.23}) at 
large $g$, as expected.  

The electron-vibron pseudopotentials are defined as
\be
u_{LM}= E^{LM}_{n+1}+E_{n-1} -2E_n   ~,
\label{5.1}
\ee
where $E_n$ are the fully relaxed ground state energies of  $n$ electrons, which igbnore
the effects
of electron-electron interactions. 
For odd values of $n$, this is an effective pairing
interaction often called ``pair binding'' in the literature\cite{ck}.
In Section \ref{semiclas} we found that for all odd $n$, the pair energies are
negative, and given by the large $g$ asymptotic expression
\be
u_{n=1,3,5}~\sim - g^2 + 1 -
{2\over 3 g^2} ~+{\cal O}(g^{-4}).
 \label{5.2} \ee
The first term is  the {\em classical energy}. The second term is due to
reduction of zero point energy along the JT manifold, since only
radial modes remain hard. This term is independent of $g$ and positive. The
last term is due to the quantum pseudo--rotator Hamiltonian, and the Berry
phases which impose a finite ground state energy associated with odd $L$ for
odd numbers of electrons. This term, although nominally small at large $g$,
becomes important at weaker coupling. If (\ref{5.2}) is extrapolated to the
weak coupling regime the last term would dominate the pair binding energy. The
exact diagonalization indeed shows a significant
enhancement of the pair binding energy over the classical value in the weak
coupling regime.

In the weak coupling limit, we can obtain analytical expressions for $u_n(g)$
for $g<<1$ by second order perturbation theory.  The unperturbed Hamiltonian is
the non interacting part  $H^0$.  The perturbing
hamiltonian is $H^{e-v}$ of Eq. (\ref{2.7}), which connects Fock states
differing by one vibron occupation. All diagonal matrix elements vanish, and
the leading order corrections to any degenerate multiplet are of order $g^2$.
These are given by diagonalization of the matrix \cite{sak},
\be
\Delta^{(2)}_{n_{ms},n_{ms}'}=
 \langle 0, n_{ms} | H^{e-v} {1 \over {E^{(0)}_a-H^0}}
H^{e-v} | 0,n_{ms}' \rangle ,
\label{5.3}
\ee
in the degenerate 0-vibrons subspace. The sum implied by the inverse operator
$(E^{(0)}_a-H^0)^{-1}$ extends just to the $N_v$=1 states. The eigenvalues of
$\Delta^{(2)}$  yield the ground state energies and splittings for different
electron fillings.  These results, for all $H_g$ and also $A_g$ modes, and
extended to the $N_v$=1 multiplet, are discussed more extensively in Ref.\cite{II}.

Here we refer only to ground state energetics. In particular, using the
perturbative expressions, we obtain, for a single $H_g$, mode the small $g$
pair binding energy
\be
{ u_{n=1,3,5}\over \hbar\omega}~= -{5\over 2} g^2~+ {\cal O}(g^4).
\label{5.6}
\ee
The origin of the 5/2 factor that characterizes the perturbative
result (\ref{5.6}) is group theoretical \cite{YB,II}.

The molecular pair binding
energy  can be considered as an effective negative-U Hubbard interaction for
the
lattice problem, provided that the Fermi energy $\epsilon_F$ is not much larger
than the JT  frequency scale $\omega$.
A mean field estimate of the transition temperature for the negative-$U$
Hubbard
model in the weak coupling regime is \cite{NegU,NegU2,ck}
\be
T_c \approx \epsilon_F
\exp \left[ \left(- N(\epsilon_F) | U |\right)^{-1} \right] .
\label{5.7}
\ee
In Refs. \cite{sch} and \cite{lannoo}, the results of Migdal--Eliashberg
approximation for the superconducting transition temperature was given. Without
the Coulomb pseudopotentials  this approach yields
\bea
T_c &\approx& \omega \exp \left[ \left(- N(\epsilon_F) |V| \right)^{-1} \right]
 \nonumber\\
V&=&-{5\over 6} g^2 ~ .
\label{5.8}
\eea
By comparing (\ref{5.6}) to (\ref{5.8}) we find a striking discrepancy between
the values of the effective pairing interaction:
\be
U~=3 V .
\ee
That is to say: in  the weak coupling regime, the correct molecular calculation
yields  a pairing interaction which is {\em three times larger} than the
results of Migdal--Eliashberg theory!

\section{TDAE-\c60:  Mott-ferromagnetism}
\label{sec4}
The fullerene compound TDAE-\c60, where \c60 is buckminsterfullerene
and TDAE is tetrakis\-(dimethylamino)\-ethylene C$_2$N$_4$\-(CH$_3$)$_8$,
exhibits ferromagnetism at  \Tc$\approx 16^\circ$K \cite{tdae-fm,tdae-fm1}.
The striking aspects of this discovery are (i) the magnitude of \Tc  --
relatively large for a material with no transition metals -- 
and (ii) its nonmetallic conductivity, suggestive of Mott-Hubbard
localization\cite{microwave}.

ESR studies \cite{tanaka} show that TDAE donates an electron to \c60.
Furthermore, no ESR signature of TDAE$^+$ is observed, suggesting that
the TDAE radical spins are somehow paired.  The monoclinic structure
makes for a relatively short inter-\c60 separation along the $c$-axis.
We are then led to consider a model of \bb{1} chains whose conduction
electrons interact via superexchange.

Superexchange in a one-band model is always antiferromagnetic,
since the intermediate state $\ket{\uar\dar}$ is a spin singlet.
The molecular degeneracy of the $t_{1u}$ \c60 LUMO leads
to interesting possibilities not realized in a orbitally nondegenerate
model.  Indeed, Seshadri \etal \cite{rao1} have discussed how
ferromagnetism naturally arises via superexchange through intermediate
states with a negative singlet-triplet splitting (\eg\ Hund's rule)
in \bb{2-}.  We introduce here what
we believe to be a `minimal model', based on the structure of
TDAE-\c60, which leads to insulating ferromagnetic behavior.

The main result in this section is the full multicomponent superexchange
Hamiltonian for the Mott insulator and the  analysis of its phase diagram.
The model is
characterized by only three pseudopoentials for the 
intermediate states of \bb{2-} in an axially symmetric crystal field.   
In the limit where all the
intermediate states are degenerate there is an accidental symmetry which leads
to the
SU(4) Heisenberg antiferromagnetic model  in its fundamental
representation. This model, coincidentally, was solved by Sutherland\cite{suth} using Bethe's {\it Ansatz}.  In the case of large singlet-triplet splitting,
the ground state is a fully polarized {\em spin} ferromagnet 
and an orbital antiferromagnet whose correlations are
given by n by Bethe's wavefunction.  
Full details of the calculations below were given in Ref. \cite{assa-dan}.

\subsection{The Hopping Model} 
We consider tight binding hopping on a lattice of \c60 molecules
with a filling of one electron per site.  
In general, a tetragonal or monoclinic
crystalline symmetry will resolve the triply degenerate $t_{1u}$ orbital
into three distinct levels. Better details could be obtained 
{\em ab-initio} once the precise structure
and orientations of the \c60 molecules are experimentally ascertained.
In our model, we shall retain only
what we believe may be the essential microscopic physics undelying
the ferromagnetism in TDAE-\c60:
(a) The hopping is quasi-one dimensional along the $c$-axis.
(b) We assume that the crystal field resolves the  
$t_{1u}$ orbital triplet into a lower doublet ($l=\pm$) and
a higher singlet $l=0$ at higher energy, as if the crystal
fields are cylindrically symmetric about an axis which pierces the
center a pentagonal face of \c60.  
(c) Hopping along the chains is assumed to preserve the orbital
magnetization $l$. 

Thus we investigate the Hamiltonian
$\cH =\cH^\parallel_{\rm hop} + \cH_{\rm hop}^\perp+\cH_{\rm ion}$,
where
\bea
\cH_{\rm hop}^\parallel & = & -\tpar\sum_{\bfi,l,\sigma}\(c\yds{l\sigma}(\bfi)
c\nds{l\sigma}(\bfi+\bfc)+{\rm H.c.}\)\nonumber\\
\cH_{\rm hop}^\perp & = & -\half\sum_{\bfi,\bdper\atop l,l',\sigma}
\tper_{ll'}(\bdper)\( c\yds{l\sigma}(\bfi)c\nds{l'\sigma}(\bfi+\bdper)+
{\rm H.c.}\)\nonumber\\
\cH_{\rm ion} & = & \sum_{\bfi,\Lambda}\ubar_\Lambda\,\ket{\Lambda(\bfi)}
\bra{\Lambda(\bfi)}\ .
\eea
Here $c\yds{l\sigma}(\bfi)$ creates, at site $\bfi$,
an electron of spin polarization $\sigma=\uparrow,\downarrow$ and
``isospin'' $l=+,-$. $\bfc$ and $\bdper$ denote
nearest neighbor lattice vectors in the $c$ direction and the
$a$-$b$ plane respectively. $t_{ll'}$ are the hopping matrix elements between
orbitals $l$ and $l'$ on  neighboring chains.

$\cH_{\rm ion}$ is the interaction Hamiltonian which
discourages multiple electron occupancy on any \c60 molecule.
It is parametrized by pseudopotentials $\ubar_\Lambda$ which correspond to the
following  \bb{2-} multiplets,
\bea
\ubar_0:&& \frac{1}{\sqrt{2}}(c\yds{+\uar}c\yds{-\dar}\!-\!c\yds{+\dar}
c\yds{-\uar})
\!\ket{0}\nonumber\\
\ubar_1:&&
c\yds{+\uar}c\yds{-\uar}\ket{0},
\frac{1}{\sqrt{2}}(c\yds{+\uar}c\yds{-\dar}+c\yds{+\dar}c\yds{-\uar})
\ket{0} ,
c\yds{+\dar}c\yds{-\dar}\ket{0}\nonumber\\
\ubar_2:&& 
c\yds{+\uar}c\yds{+\dar}\ket{0}, 
c\yds{-\uar}c\yds{-\dar}\ket{0} 
\eea
The relations between $\ubar_\Lambda$ and the isotropic pseudopotentials $u_L$
of angular momenta $L$  are: $\ubar_1= u_1$, $\ubar_2=u_2$, but $\ubar_0=
{2\over 3} u_0 + {1\over 3} u_2$ due to projecting out the ``$l=0$''
orbital state.  Thus, while in an isotropic environment there might be pair binding  ($u_0 < 0$) due to electron-electron \cite{ck} and electron-vibron \cite{amt} interactions, it does not preclude
a repulsive $\ubar_0 > 0$ in the monoclinic crystal field environment. 
This may help to explain why TDAE-\c60 is not a CDW, nor
a superconductor as is \a3c60.

\subsection{Multicomponent Superexchange Hamiltonian}
Experiments have shown that TDAE-\c60 is insulating at low temperatures,
consistent with the existence of a gap to charge fluctuations
(i.e. all $\ubar_\Lambda>0$) \cite{microwave}.
The low-lying excitations can be described by a superexchange
Hamiltonian, formally obtained as a second order expansion in small
$\tpar/\ubar$. Since charge excitations are gapped, a renormalized
version of the superexchange Hamiltonian is expected to describe the low
energy excitations also for $\tpar/\ubar\gtwid 1$. 

The zeroth order states of the superexchange Hamiltonian are four singly
occupied states enumerated by $l,\sigma$. The  operators 
which act on these states can be represented by 
spin  operators,  
$S^\mu_\bfi=\half\sum_{l,\sigma,\sigma'} c\yds{l\sigma}(\bfi)\,
\tau^\mu_{\sigma\sigma'}\,c\nds{l\sigma'}(\bfi)$ 
and  ``isospin'' operators, 
$I^\nu_\bfi=\half\sum_{l,l',\sigma} c\yds{l\sigma}(\bfi)\,
\tau^\nu_{ll'}\,c\nds{l'\sigma}(\bfi)\ $
where $\bftau$ are the Pauli matrices.  
Taking into account the constraint  
$\sum_\alpha c\yds{\alpha}c\nds{\alpha}=1$, the 15 independent elements of
the SU(4) generators  
$S^\alpha_\beta= c\yds{\alpha}c\nds{\beta}$ can be expressed in terms of
the 15 operators
$\{S^\mu,I^\nu,S^\mu I^\nu\}$.

For simplicity we  consider the purely one-dimensional limit where
$t^\perp_{ll'}=0$.  There are three superexchange constants defined as
\be
J_M\equiv {2(\tpar)^2\over\ubar_M}\ \,\qquad M=0,1,2\ ,
\ee
A straigthforward, though cumbersome, leads to an effective Hamiltonian given
by
\bea
\cHt&=&\sum_n\(A\,\bfS_n\cdot\bfS_{n+1}+B\,\bfI_n\cdot\bfI_{n+1}
+C\, I^z_n I^z_{n+1}+\right.\label{sex1}\\
&&\left.+D\,\bfS_n\cdot\bfS_{n+1}\,\bfI_n\cdot\bfI_{n+1}
+E\,\bfS_n\cdot\bfS_{n+1}\,I^z_n I^z_{n+1} +F\)\ ,\nonumber
\eea
where
\bea
A =-\half J_1+J_2+\half J_0,&~~~~ 
&B=  \frac{3}{2} J_1-\half J_0,\nonumber\\
C=  J_0-J_2 ,&~~~~ 
&D =  2 J_1+2 J_0,\nonumber\\
E=  4 J_2 - 4 J_0 ,&~~~~  
&F =  -\frac{3}{8} J_1 - \fourth J_2 - \frac{1}{8} J_0\ .
\eea
This model possesses a global SU(2)$\times$U(1) symmetry, \ie\ $\cHt$
commutes with $\sum_n \bfS_n$ and with $\sum_n I^z_n$.  Enlarged symmetries
occur when $J_0=J_2$, where the symmetry group is SU(2)$\times$SU(2), and
when $J_0=J_1=J_2$, where the symmetry group is SU(4).

{\em SU(4) Point --} At the point $\ubar_1=\ubar_2=\ubar_0\equiv\ubar$
(\ref{sex1}) acquires full SU(4) symmetry. For each $c$-chain the Hamiltonian is
\be
\cH_{SU(4)}=J\sum_n \sum_{\alpha,\beta}S^\alpha_\beta(n)S^\beta_\alpha(n+1)
\ ,
\ee
where $J=2(\tpar)^2/\ubar$.  

The SU($P$) Heisenberg antiferromagnet in the fundamental
representation has been solved by Sutherland for general $P$ using
Bethe's {\it Ansatz\/} \cite{suth}.
This model exhibits $P-1$ gapless elementary
excitation branches.  We presume, based on what happens in the SU(2)
model \cite{fata}, that for a chain of $N$ sites where $N$ is an
integer multiple of $P$, the ground state is an SU($P$) singlet and the
low-lying excitations transform according either to the singlet or
the adjoint representation. This is essentially what happens in the
fermion mean field (large $P$) theory of the
SU($P$) antiferromagnet \cite{bza,AA,RS}. The mean field has four degenerate
quarter-filled ($\kF=\fourth\pi$) bands for $P=4$.  Although there it
has no true long-ranged order, the spin and isospin susceptibilities
diverges at the nesting wavevector $2\kF=\half\pi$, which describes a
commensurate spin density wave of period four.  The period four arises
because the spin chain is in its fundamental representation, and by
`4-ality' one needs four sites to make a singlet \cite{comm3}. The mean
field theory also predicts a constant uniform (Pauli) susceptibility, and
a linear specific heat as in a Fermi liquid\cite{AA}.  

{\em Ferro-Antiferromagnetic points --}
Along the surface $\ubar_2=\ubar_0$, our Hamiltonian possesses an
SU(2)$\times$SU(2) symmetry.  There are then two special limits in
which we can determine the exact ground state.
(i) The ``$\rmF\times\rmA$ model''
at $\ubar_0\to\infty$, with $J_\parallel= 2 (\tpar)^2/\ubar_1$,
\be 
\cH_{\rmF\times\rmA}=-{4(\tpar)^2\over\ubar_1}\sum_n
(\bfS_n\cdot\bfS_{n+1}+\frac{3}{4})
(\fourth - \bfI_n\cdot\bfI_{n+1})\label{HamFA} 
\ee
where the interactions are ferromagnetic in
the spin channel and antiferromagnetic in the isospin channel, and
(ii) the ``$\rmA\times\rmF$  model'' for 
$\ubar_1\to\infty$, with $J_\parallel= 2 (\tpar)^2/ \ubar_0$, and the
roles of $\bfI$ and $\bfS$ interchanged. 

It is possible to prove that the ground state of $\cH_{\rmF\times\rmA }$
is the fully polarized ferromagnet $\ket{\rmF}_S$ for the spin variables,
and Bethe's ground of the spin-half antiferromagnet for the
isospin variables \ie
\be
\Psi^{\rmF\times\rmA }_0 =  \ket{\rmF}_S\otimes\ket{{\rm Bethe}}_I\ .
\label{prodwf}
\ee
A corresponding result holds for $\cH_{\rmA\times\rmF}$, with spin and isospin
variables exchanged.
Due to the $SU(2)\otimes SU(2)$ symmetry the total spin $S_{\rm tot}$, total
isospin $I_{\rm tot}$,
and their polarizations along the $\zhat$ axis ($M_S$ and $M_I$,
respectively) are good quantum numbers.
Following Lieb and Mattis' proof of the Marshall theorem for the
Heisenberg model\cite{LM}, we perform  a $\pi$ rotation about the
$\zhat$ axis of the isospin operators on odd-numbered sites. 
The Hamiltonian transforms into  a non-positive (`negative semidefinite')
operator in the product Ising basis
\bea 
\cH_{\rmF\times\rmA }&\to&
J\sum_n(I^z_n I^z_{n+1}-\half I^+_n I^-_{n+1}-\half I^-_n I^+_{n+1}
-\fourth)\nonumber\\
&&\qquad\times 
(\bfS_n \cdot \bfS_{n+1} +\frac{3}{4})\equiv\cH'_{\rmF\times\rmA }
\eea
The accessibilty of all states within a given
magnetization sector by repeated application of the Hamiltonian,
implies (see Ref. \cite{LM}) that the ground state of
$\cH'_{\rmF\times\rmA }$ in the sector 
$(M_S,M_I)=(0,0)$ can be chosen to be positive definite in the
sublattice-rotated Ising basis, \ie\ it obeys Marshall's sign rule.
Since the same Marshall signs hold for the state on the right hand side
of  Eq. \ref{prodwf},
which has $S_{\rm tot}=\half N $, and $I_{\rm tot}=0$,
the two sides of Eq. \ref{prodwf} have finite overlap hence the same
$S_{\rm tot}$ and $I_{\rm tot}$.
We are free to choose $M_S=\half N$ as a representative of the ground
state manifold.
Note that $\ket{\Psi^{\rmF\times\rmA  }_0}$ is indeed an eigenstate of the
spin triplet
projection operator $(\bfS_n\cdot\bfS_{n+1}+\frac{3}{4})$ with
eigenvalue one.
It follows from Eq. \ref{HamFA} that the isospin part of the wavefunction
is the ground state of the spin-half
antiferromagnetic Heisenberg chain, given by Bethe's {\it Ansatz}.

Exact excitations of $\cH_{\rmF\times\rmA}$
within the isospin sector (retaining full spin
polarization) with dispersion $\half\pi J|\sin k|$ can be constructed
as in Refs. \cite{fata,dcp}.  The gapless ferromagnetic magnons,
which exist due to Goldstone's theorem, can be approximated within the
Single Mode Approximation (SMA): $\ket{k}\equiv
S^-_k\ket{\Psi_0^{\rmF\times\rmA}}$.  The trial state dispersion is
\be
\omega( k) \le  2\ln( 2)J 
 (1-\cos k )
\ee 
from which we see that the ferromagnon bandwidth is decreased due to the
antiferromagnetic nearest-neighbor isospin correlations, \ie\ 
$\langle \fourth - \bI_n \cdot \bI_{n+1} \rangle = \ln(2)$.

\subsection{Classical Phase Diagram}
The ground state depends on the dimensionless  ratios 
${\bar u}_0/{\bar u_1}$ and  ${\bar u}_2/{\bar u_1}$. The classical
approximation (justified at  $S,I>>1$) is given by minimizing the bond
energies of Eq. (\ref{sex1}) as function of vectors  $\bfS_\bfi $ and
$\bfI_\bfi$ of magnitude $\half$.  The results are plotted in
Fig. \ref{fig-phase}).

It is interesting to note that the
SU(4) symmetry point is at the border of 4 distinct ordered phases of
different symmetries, where the energy is degenerate along the lines
$\langle \bI_n\cdot\bI_{n+1}\rangle=-{1\over 4}$ and
$\langle \bS_n\cdot\bS_{n+1}\rangle={1\over 4}$.
The large degeneracy of the classical SU(4) model is reduced 
by quantum fluctuations.

The class\-ical reg\-ime of Heis\-enberg spin\---ferro\-mag\-net\-ism and
iso\-spin-anti\-ferro\-magnetism extends throughout $\ubar_0,\ubar_2 > \ubar_1$,
although quantum fluctuations break the SU(2) isospin symmetry away from
the isotropic line $\ubar_0=\ubar_2$ (marked as a dashed line in
Fig. \ref{fig-phase}).

{\em 3D Ordering in the $\rmF\times\rmA$ Model --}
As shown by Scalapino \etal \cite{sip}, one can treat
the interchain interactions by mean field theory and thereby derive an
expression for the full susceptibility $\xhi_{ab}(\bfq_\perp,q_z,\omega)$
in terms of $\xhi^{\rm 1D}_{ab}(q_z,\omega)$, the susceptibility for
the one-dimensional chains.  The general result is
\be
\xhi(\bfq_\perp,q_z,\omega)=\sbl \mbox{\bf 1}-J_\perp(\bfq_\perp)\,\xhi^{\rm 1D}
(q_z,\omega)\sbr^{-1}\xhi^{1D}(q_z,\omega)\ ,
\label{susc}
\ee
where $J_\perp(\bfq_\perp)=\sum_{\bdper} J_\perp(\bdper)\,
e^{-i\bfq_\perp\cdot\bdper}$
is the spatial Fourier transform of the interchain
coupling matrix.  (Note that the quantities $\xhi$, $J_\perp$, and 
$\xhi^{\rm 1D}$ in Eq. \ref{susc} are matrices.)  This approximation
also may be employed at finite temperature.

Consider now the F$\times$A model discussed above.  At finite
temperature $T$, long-ranged ferromagnetic order is destroyed and
the global SU(2)$\times$SU(2) symmetry is restored.  
The uniform susceptibility of the ferromagnetic chain is given by 
\be
\xhi_\rmF'(0,0;T)={J_\parallel\over 24 T^2}+\ldots\ ,
\ee
as was first derived by Takahashi in Ref. \cite{tak}
(see also Refs. \cite{AA,assa}).

For the antiferromagnetic susceptibility, we appeal to the bosonization
results of Schulz and of Eggert and Affleck \cite{ian2}, who have
computed the dynamic susceptibility of the $S=\half$ antiferromagnetic  Heisenberg
chain. Performing a Fourier transform of their result and taking the
low frequency and wavevector limit near the antiferromagnetic point
we obtain the {\it staggered} isospin susceptibility
\be
\xhi_\rmA  \approx {a_0^2\over\pi T}
\ee
where $a_0\simeq 4.44$. 

For mixed interchain coupling operators  
\eg\  $\cO=S^x I^y$, we may use the assumed independence of
ferromagnetic and antiferromagnetic magnons to obtain
at low
temperatures $\xhi'_{\rmF\rmA}(\pi,0;T)\sim (J_\parallel T)^{-1/2}$,
which diverges even more slowly than $\xhi_\rmA$ in the $T\to 0$ limit.

The interchain interaction is given by
$J_\perp=J_\parallel((t^\perp_1)^2+(t^\perp_2)^2)/4(t^\parallel)^2$,
where $t^\perp_{1},t^\perp_{2}$
are the transverse hopping integrals (see Ref. \cite{assa-dan}).
Thus, as the temperature is lowered, a transition from paramagnetic to
ferromagnetic state should set in when $J_\perp\xhi'_\rmF=1$.
This yields $T_\rmC\simeq\sqrt{J_\parallel J_\perp/24}$.  The relation
$T_\rmC\propto\sqrt{J_\parallel J_\perp}$ was also found by
Scalapino \etal\ (ref. \cite{sip}) in their studies of anisotropic
Heisenberg magnets.  It is conceivable that at still
lower temperatures a N{\'e}el ordering of the isospin variables occurs at
a N{\'e}el temperature $T_\rmN\simeq 3 a_0^2 J_\perp/\pi$.

\subsection{Experimental notes}  
(a) The lower isospin
transition, to our knowledge, has not been resolved experimentally. Perhaps 
it is not very well separated from the ferromagnetic transition which
would help explain the mysterious excessive entropy of transition found by
Ref.\cite{tanaka}. 

(b) Alternatively, the isospin ordering might be preempted by a
{\em isospin-Peierls}
ordering (orbital dimerization) aided by the electron-phonon coupling. In
that case, a signature for the isospin-Peierls effect should be present
in X-ray scattering or in the phonon spectrum. 

(c) The role of a possible 
orientational disordered ground state \cite{brooks} has not been considered here
although it might help explain the observed weak ferromagnetism
\cite{tdae-fm}. In addition,
Bloch's $T^{3/2}$ temperature dependence of the ordered moment found in
Ref. \cite{tdae-fm1} which holds upto $T\approx T_\rmC$ is hard to reconcile 
with  quasi-one dimensionality where $J_\perp << T_\rmC$. 

In summary, 
this section has described a model of 
quasi-one dimensional interacting electrons with doubly degenerate orbitals
motivated by the structure of TDAE-\c60. At occupancy of one
electron per site, we obtain a Mott-insulator with  muticomponent
superexchange between
spins and isospins at neighboring sites. At special values of the
interactions we identify exactly solvable points, including the SU(4)
antiferromagnet, and spin-ferromagnet, isospin-antiferromagnet limit.
The classical ground state diagram also
contains a large region of spin ferromagnetism and orbital
antiferromagnetism which we believe is relevant for  TDAE-\c60.
A mean field analysis of the interchain coupling in this regime
predicts two transition temperatures: ferromagnetic spin ordering at
$T_\rmC\propto \sqrt{J_\parallel J_\perp}$, and  orbital (isospin)
antiferromagnetic ordering at $T_\rmN\propto J_\perp $. This lower
transition, to our knowledge, has not yet been resolved experimentally.

\section{Concluding remark}
In this review we have seen that it is possible to understand some
of the  unusual electronic properties of
buckminster\-fullerene compounds by relating them to the high
molecular symmetry, and local degeneracies of the partially
filled $t_{1u}$ orbitals. We have investigated in detail some simplified
models with local degeneracies, which exhibit enhanced superconductivity, and ferromagnetism, as well as other possibile phases not yet observed experimentally.
This suggests that continual experimenting with the family of fullerene compounds 
would most probably  produce further surprises. 

\subsubsection*{Acknowledgements}
I am greatly  indebted to my collaborators Daniel Arovas, Nick Manini, Ganpathy Murthy and  Erio Tosatti. Useful discussions with Yossi Avron, Mary O'Brien,
Seb Doniach, Olle Gunnarson and Steve Kivelson are gratefully acknowledged.
This work was supported by grants
from the  Israeli Academy of Sciences and the Fund for Promotion of
Research at Technion.


\begin{table}
\caption{Semiclassical ground state distortions and energies for a single $H_g$
coupled mode of frequency $\omega$. $n$ is the electron number, $S$ is the
total spin, ${\bar z}_n,{\bar r}_n$ are the JT distortions, ${\bar n}_i$ is the
occupation of orbital $i$, $E_n$ is the ground state energy and, and $u_n$ is
the pair energy (Eq.(\protect\ref{5.1})). Energies are calculated for strong coupling
to order $g^{-2}$.}
\vskip 0.3in
\begin{tabular}{|c|c|l|c|c|c|} \hline
$n$&$S$& $({\bar z}_n,{\bar r}_n)$ & $({\bar n}_1,{\bar n}_2,{\bar n}_3)$ &
$ E_n/(\hbar\omega)$ &$U_n/(\hbar \omega)$ \\ \hline
0 &0		& $(0 ,0)$ & (0,0,0)&  ${5\over 2}$&   \\
1 &$\frac{1}{2}$& $(g ,0)$  & (0,0,1)&  $-\half g^2 +{3\over 2}+{1\over
3g^2} $&   $  -g^2+1-{2\over 3g^2}  $
\\ 2 &0		& $(2g ,0)$ & (0,0,2) &
$-2g^2+{3\over 2}  $&  \\
3
&$\frac{1}{2}$& $({3\over 2} g ,  {\sqrt{3}\over 2} g )$  & (1,0,2) &
$ -{3\over 2}
g^2+1+{1\over 3g^2} $& $ -g^2+1-{2\over 3g^2}  $       \\
 4 &0		& $(-2g
,0)$  & (2,2,0)  &   $-2g^2+{3\over 2}  $&  \\
5
&$\frac{1}{2}$& $(-g ,0)$
 & (2,2,1)&  $ -\half g^2 +{3\over 2}+{1\over
3g^2}    $& $   -g^2+1-{2\over 3g^2}     $  \\
6 &0		& $(0 ,0)$ & (2,2,2)& ${5\over 2}$ &  \\ \hline
\end{tabular}
\end{table}

\begin{figure}
\caption{A polar representation of the Jahn-Teller distortions
$u^{JT}({\tilde \theta},{\tilde \phi})$, Eq.(\protect\ref{3.6}).
The distortion is measured relative to a sphere.
(a) The unimodal distortion for the
ground states of $n=1,2,4,5$ electrons (b) The bimodal distortion
for $n=3$ electrons.}
\label{dist}
\end{figure}

\begin{figure}
\vskip 2in
\caption{TDAE-\c60: Classical ground state phase diagram}
\label{fig-phase}
\end{figure}

\end{document}